\documentclass[conference]{IEEEtran}

\usepackage{xcolor,soul,framed} 

\colorlet{shadecolor}{yellow}
\usepackage[pdftex]{graphicx}
\graphicspath{{../pdf/}{../jpeg/}}
\DeclareGraphicsExtensions{.pdf,.jpeg,.png}

\usepackage[cmex10]{amsmath}
\usepackage{array}
\usepackage{mdwmath}
\usepackage{mdwtab}
\usepackage{eqparbox}
\usepackage{url}
\usepackage{balance}
\usepackage[noadjust]{cite}

\begin{document}
   \title{Intelligent Transportation Systems' Orchestration: Lessons Learned \& Potential Opportunities}

 \author{\IEEEauthorblockN{Abdallah Moubayed\IEEEauthorrefmark{1}, Abdallah Shami\IEEEauthorrefmark{1}, and Abbas Ibrahim\IEEEauthorrefmark{2}}
\IEEEauthorblockA{\IEEEauthorrefmark{1} Western University, London, Ontario, Canada;\\ 
e-mails: \{amoubaye, abdallah.shami\}@uwo.ca\\
}
\IEEEauthorblockA{\IEEEauthorrefmark{2}  ISSAE-Cnam Liban affiliated to CNAM Paris;\\
e-mail: aibrahim@cnam.fr}
 }

\markboth{IEEE Transactions on Intelligent Transportation Systems}{Moubayed \MakeLowercase{\textit{et al.}}: }

\maketitle

\begin{abstract}
The growing deployment efforts of 5G networks globally has led to the acceleration of the businesses/services' digital transformation. This growth has led to the need for new communication technologies that will promote this transformation. 6G is being proposed as the set of technologies and architectures that will achieve this target. Among the main use cases that have emerged for 5G networks and will continue to play a pivotal role in 6G networks is that of Intelligent Transportation Systems (ITSs). With all the projected benefits of developing and deploying efficient and effective ITSs comes a group of unique challenges that need to be addressed. One prominent challenge is ITS orchestration due to the various supporting technologies and heterogeneous networks used to offer the desired ITS applications/services. To that end, this paper focuses on the ITS orchestration challenge in detail by highlighting the related previous works from the literature and listing the lessons learned from current ITS deployment orchestration efforts. It also presents multiple potential data-driven research opportunities in which paradigms such as reinforcement learning and federated learning can be deployed to offer effective and efficient ITS orchestration.
\end{abstract}


%
\IEEEpeerreviewmaketitle


\section{Introduction}\label{intro}
\indent The continually growing deployment efforts of 5G networks globally has led to the acceleration of the businesses/services' digital transformation \cite{5g_devices1}. This transformation has been further amplified by the pandemic by forcing a faster transition into the digital world \cite{6G_ref1}. Accordingly, the market size growth for 5G services is steadily growing with estimates that it will reach  665 billion USD by 2028  with massive Machine-Type communication applications having the majority share in terms of growth rate \cite{5G_market_size}. \\
\indent This growth has led to the need for new communication technologies that will promote this transformation while also enabling the sustainability of the systems available. Accordingly, 6G is being proposed as the set of technologies and architectures that will achieve this target by allowing users and businesses to interact and control the digital and virtual worlds that are decoupled from their physical location. Because of the huge promise that 6G networks offer, it is projected that the that the global market size for 6G will reach 1,773 billion USD by the year 2035 after an anticipated launch date of 2030 \cite{6G_market_size}.\\
\indent Among the main use cases that have emerged for 5G networks and will continue to play a pivotal role in 6G networks is that of Intelligent Transportation Systems (ITSs). The development of efficient ITSs has attracted attention from a multitude of stakeholders due to the various benefits it offers ranging from traffic-related accident reduction to the cost-effective management of vehicular fleets \cite{v2x_access_survey1,v2x_access_survey_moubayed}. \\
\indent With all the projected benefits of developing and deploying efficient and effective ITSs comes a group of unique challenges that need to be addressed. One prominent challenge is ITS orchestration \cite{5G_ITS_challenges}.  Orchestration can be defined in a multitude of ways. One such definition is that given by the Open Network Foundation which refers to orchestration as ``\textit{the selection of resources to satisfy service demands in an optimal way, where the available resources, the service demands and the optimization criteria are all subject to change}'' \cite{orchestration_def1}. A similar definition is given by the European Telecommunications Standards Institute which states that orchestration is ``\textit{the coordination of the resources and networks needed to set up cloud-based services and applications}'' \cite{orchestration_def2}. These definitions are particularly relevant to the ITS case due to the various supporting technologies (e.g. softwarization, virtualization, etc.) and heterogeneous networks used to offer the desired applications and services. Hence, managing and coordinating these networks by allocating the appropriate available resources is a challenging task due to the high mobility of the users and the stringent performance requirements of the corresponding ITS services and applications. \\
\indent To that end, this paper focuses on the ITS orchestration challenge in detail. It also presents some of the data-driven research opportunities that can help address this challenge. Accordingly, this paper: 
\begin{itemize}
	\item Provides a brief background on ITS systems and the corresponding set of enabling technologies.
	\item Details the challenge ITS systems' orchestration by highlighting the related previous works from the literature.
	\item Lists the lessons learned from current ITS deployment orchestration efforts.
	\item Presents different data-driven research opportunities and approaches to address this challenge.
\end{itemize}\mbox{}\\
\indent The remainder of this paper is organized as follows: 
Section \ref{ITS_orch_related} describes the ITS orchestration challenge, presents some of the previous works from the literature tackling it, 
and lists the learned lessons from current deployments. 
Section \ref{opportunities} describes several data-driven research opportunities that can be adopted to tackle the ITS orchestration challenge. Section \ref{conc} concludes the paper.
\section{ITS: Background \& Enabling Technologies}\label{ITS_background}
\subsection{ITS Background:}
\indent Modern ITS systems can be characterized by the communication modes they support and the set of applications and services they offer \cite{ITS_background_moubayed}. On the one hand, the communication modes (commonly referred to as V2X communication modes) define how the entities within the ITS communicate and coordinate among each other to ensure the efficiency, effectiveness, and safety of the system \cite{ITS_background_moubayed}. On the other hand, the applications and services that ITS systems offer rely heavily on these communication modes as they facilitate the autonomy of the system \cite{ITS_background_moubayed}.\\
\indent Starting with the communication modes, ITS systems typically support four main modes as follows \cite{v2x_communication_modes}:
\begin{enumerate}
    \item Vehicle-to-network (V2N) Communication: This mode focuses on the communication between a vehicle and a V2X application server \cite{C-V2X1,C-V2X2}.
    \item Vehicle-to-vehicle (V2V) Communication: This mode focuses on the  direct communication between two vehicles \cite{v2v_def}. 
    \item Vehicle-to-infrastructure (V2I) Communication: This mode focuses on the communication between a vehicle and roadside infrastructure such as road-side units (RSUs) \cite{v2i_def}. 
    \item Vehicle-to-pedestrian (V2P) Communication: This mode focuses on communication between vehicles and vulnerable road users (VRUs) such as pedestrians, wheelchair users, bikers, and motorcyclists \cite{v2p_def}.
\end{enumerate}

\indent In terms of the applications and services offered, ITS systems focus on a set of main categories as follows:
\begin{enumerate}
    \item Traffic Safety: This category focuses on the set of applications and services that ensure the safety of the different entities within the ITS \cite{traffic_safety1}. 
    \item Traffic Efficiency: This category focuses on applications/services that aim at enhancing the moving fluency and smoothness of traveling vehicles on the roads \cite{traffic_efficiency1}. 
    \item Autonomous/Cooperative Driving: This category focuses on the applications/services that allow vehicles to drive in an autonomous manner through cooperation \cite{autonomous_driving1}. 
    \item Infotainment: This category focuses on the applications/services that provide entertainment as well as general information not related to driving \cite{infotainment1}.
\end{enumerate}
\subsection{Enabling Technologies:}
\indent In order to support the aforementioned communication modes and offer the different applications/services, a set of enabling technologies have been proposed for ITSs.
In what follows, a brief overview of these technologies is provided.
\subsubsection{Heterogeneous Access Technologies}
\indent The first set of enabling technologies for ITSs is the heterogeneous access technologies. As mentioned earlier, different access technologies have been proposed in the literature that are used in different ITS communication modes and facilitate different ITS applications/services. This includes:
\begin{itemize}
    \item Celluar V2X: The first type of access technology used is cellular networks such as LTE, LTE-Advanced (LTE-A), or 5G  (including mmWave and 5G-NR) \cite{cellular_v2x_0,TCOMAnas}. These technologies are often used for V2N communication. The main merit is the ease of resource allocation due to the centralized coordination available within such networks.
    \item DSRC: The second type of access technology is DSRC, which is a modification of the WiFi standard \cite{dsrc_v2x}. This technology is often used for V2V communication by allowing vehicles to form ad-hoc networks that dynamically change. 
    \item Bluetooth: A third technology proposed to enable effective communication in ITSs is Bluetooth, particularly the Bluetooth low energy (BLE) 5.0 standard \cite{bluetooth_v2x}. The main advantage of Bluetooth is the fact that it is a widely accepted technology with high market penetration. Moreover it does not need infrastructure and hence can be effective for V2V communication \cite{bluetooth_v2x}.
    \item VLC: Another access technology that has been gaining recent interest for ITSs is visible light communication \cite{vlc_v2x}. In this case, visible light frequencies are used when transmitting and receiving data \cite{vlc_v2x}. The attractiveness of adopting VLC in ITSs lies in the fact that it can offer high data rates and does not cause interference to the over-crowded lower-frequency spectrum. Hence, it has been proposed for variety of ITS applications such as traffic safety and infotainment \cite{vlc_v2x}.
\end{itemize}
\subsubsection{Softwarization Technologies}
\indent Softwarization technologies are considered to be key enablers of ITSs. In particular, Software-Defined Networks (SDN) represent a pillar of current and future ITS deployments \cite{sdn_for_v2x_2}. This is because SDN concept can offer flexibility, scalability, and robustness to the ITS system by decoupling the control plane from the data plane \cite{sdn_for_v2x_1,sdn_for_v2x_2}. Moreover, all the entities within the ITS (such as vehicles, pedestrians, RSUs, etc.) can act as SDN switches \cite{sdn_for_v2x_1}. Moreover, SDN can enable multi-tenancy. In this case, different stakeholders (such as government agencies or vehicle manufacturers) can share the vehicular network infrastructure and offer different applications/services \cite{sdn_for_v2x_2}.
\subsubsection{Virtualization Technologies}
\indent Virtualization technologies are another set of key enablers of modern ITSs. More specifically, paradigms such as cloud and edge computing as well as network function virtualization (NFV) continue to play a pivotal role in deployed ITSs.\\
\indent Cloud and edge computing have been heavily proposed in the literature as part of ITSs. This is because they complement each other and allow ITSs to support various applications and services. Cloud computing offer a set of virtual and dynamically scalable computing, storage, and memory resources \cite{cloud_for_v2x, AbuSharkhIWCMC}. These resources are typically used to store, respond, and analyze queries corresponding to the different vehicles within the ITS \cite{cloud_for_v2x}. In contrast, edge computing provides distributed computing and storage in closer proximity to the users resulting in lower latencies. Thus, edge computing has been proposed for real-time traffic surveillance, adaptive video streaming, and traffic safety  \cite{edge_for_v2x1,NFV_SFC}.\\
\indent NFV is a key enabling technology of ITSs. This is because it allows software-based applications and functions to run on standard general-purpose servers instead of dedicated hardware \cite{NFV_def}. In turn, this provides more flexibility and scalability as well as reduces the capital and operational expenditures \cite{NFV_SFC}. Accordingly, NFV has been proposed in ITSs to offer and support new services for intelligent on-board systems (IOSs) \cite{NFV_for_V2X, Shaer_V2X}. In this case, IOSs are virtualized as software-based applications, making them easier to update and faster to run \cite{NFV_for_V2X}.
\section{ITS Orchestration: Description, Related Works, \& Lessons Learned}\label{ITS_orch_related}
\subsection{Description:}
\indent Orchestration of ITSs in an effective and efficient manner is a prime concern and challenge for the different stakeholders involved. This is caused and exacerbated by multiple factors. One crucial factor is the heterogeneous nature of ITSs and their corresponding network architecture \cite{ITS_orch_chall1}. As illustrated earlier, ITSs typically include multiple access technologies ranging from cellular to short-range communication technologies to enable the different communication modes required. As such, to simultaneously coordinate and allocate the resources of the various technologies is a challenging task, especially given the stringent performance requirements of some of the ITS applications/services \cite{ITS_orch_chall1}.\\
\indent A second equally important factor is the ITS scalability \cite{ITS_orch_chall2}. Scalability has a direct impact on the effectiveness and efficiency of ITS orchestration solutions. This is because ITSs are expected to serve thousands and millions of active connected and autonomous vehicles, particularly during rush hours along major roads in large cities \cite{ITS_orch_chall2}. This is further compounded by the number of planned and unplanned data packets exchanged between the different entities (vehicles, RSUs, VRUs, etc.) within the ITSs. This adds a further layer of orchestration complexity that needs to be addressed.
\subsection{Related Works:}
\indent To tackle the ITS orchestration problem, several previous works from the literature have proposed various potential solutions  \cite{ITS_orch_related0,ITS_orch_related1,ITS_orch_related2,ITS_orch_related3,ITS_orch_related4}. These solutions can be classified into one of four main categories, namely optimization modelling-based, heuristic-based, metaheuristic-based, or machine learning (ML)-based. \\
\indent Moubayed \textit{et al.} focused on the problem of cost-optimal ITS service placement in distributed cloud/edge computing environment \cite{ITS_orch_related0}. The goal was to determine a service placement that would reduce the deployment cost while still meeting the services' delay requirements. The authors formulated the problem as an integer linear programming problem. Moreover, the authors developed a low-complexity heuristic solution that achieved close to optimal performance \cite{ITS_orch_related0}. \\
\indent Khan \textit{et al.} focus on transmit rate control of the vehicles to ensure that multiple ITS services are supported \cite{ITS_orch_related1}. To that end, the authors proposed a heuristic orchestration solution consisting of two complementary mechanisms. The first is a multi-factor prioritization function that calculates the priority of the service based on rank, usefulness, and urgency. The second mechanism is a budgetary scheduler that uses the priority scores of the different services to dedicate portions of the transmit time to each services \cite{ITS_orch_related1}.\\
\indent Pokhrel proposed a modified version of SDN-enabled ITSs \cite{ITS_orch_related2}. In his work, the author proposed transferring some of the SDN controller functionalities from the edge nodes to the vehicles themselves. The goal was to provide an automated orchestration framework that would improve the interoperability and security of the SDN-enabled ITS. The proposed framework integrates scalability and security by design while simultaneously incorporating intent-based data networking \cite{ITS_orch_related2}.\\
\indent Dalgkitsis \textit{et al.} focused on the ITS service migration problem in vehicular networks \cite{ITS_orch_related3}.To that end, the authors proposed the use of genetic algorithm (a metaheuristic evolutionary algorithm) to solve the aforementioned problem. The goal of the proposed service orchestration framework is to retain the average service latency minimal by offering personalized service migration, while tightly packing as many services as possible in the edge of the network, for maximizing resource utilization \cite{ITS_orch_related3}.\\
\indent Yuan \textit{et al.} proposed the use of deep reinforcement learning method for cross-domain resource orchestration in  edge-enabled ITSs \cite{ITS_orch_related4}. More specifically, the authors proposed a multi-agent-based framework in which the agents utilized swarm intelligence to simultaneously optimize the traffic flow and information flow for their corresponding vehicles. This framework considered not only the information domain (containing the computing and communication resources), but also the transportation domain (containing the micro and macro resources of the intersections and roads) \cite{ITS_orch_related4}.
\subsection{Lessons Learned:}
\indent Within the context of ITSs, there are also a list of lessons learned based on the demos, trials, and test deployments performed as part of multiple projects conducted by various organizations including 5G-MOBIX, 5G-CARMEN, and 5G-CroCo \cite{5g_LL_ITS1,5g_LL_ITS2,5g_LL_ITS3,5g_LL_ITS4}. Some of the lessons learned include:
\begin{itemize}
    \item Performance limitations influenced by the OBU design \cite{5g_LL_ITS1}: The OBU design has a significant impact on the overall performance. This is due to the fact that the performance of 5G chipset is reliant on both the integrated hardware and software used. Hence, there is a need to investigate the most suitable software/hardware combination for autonomous vehicles specifically and ITSs in general \cite{5g_LL_ITS1}.
    \item Significantly attenuated signal strength observed at the OBU for internal antennas \cite{5g_LL_ITS1}: When utilizing internal antennas, the received signal strength was observed to be attenuated. Hence, there is a need to develop external antennas to improve the signal strength \cite{5g_LL_ITS1}.
    \item Accurate synchronization of connectivity equipment across all involved entities \cite{5g_LL_ITS2}: To ensure the performance is maximized, there is a need for accurate synchronization across the network core an edge as well as the road side units and vehicles. This will provide insights into the actual performance limits of the supporting networks and technologies \cite{5g_LL_ITS2}.
    \item Integration, testing and piloting requires a significant time commitment \cite{5g_LL_ITS3}: A layer testing approach is required for testing and piloting of 5G for ITSs. This approach first should focus on connectivity first, followed by the applications, and finally the vehicles \cite{5g_LL_ITS3}. Accordingly, operators should not attempt to do everything simultaneously or all combined \cite{5g_LL_ITS3}.
\end{itemize}
  
\section{Data Driven Research Opportunities For ITS Orchestration}\label{opportunities}
\indent Although different ML models and paradigms have been previously proposed in the literature to assist in ITS orchestration, there still exists further data-driven research opportunities worth exploring. In what follows, three main opportunities are presented and discussed.
\subsection{ITS Orchestration Using Reinforcement Learning:}
\begin{figure}[!htb]
	\centering
	\includegraphics[scale=.32,trim=0cm 0.5cm 0cm 1cm]{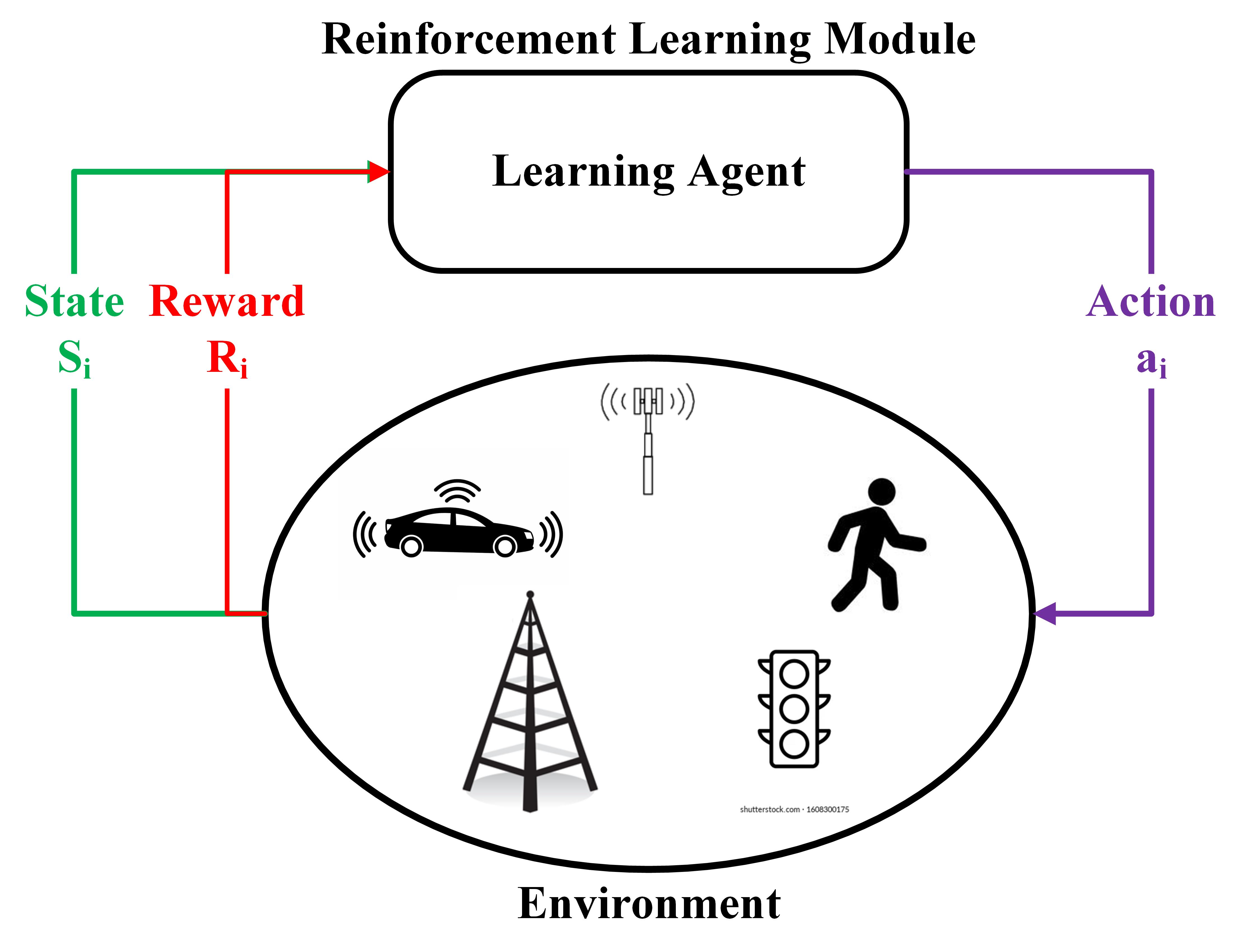}
	\caption{Potential RL-enabled ITS Orchestration Architecture}
	\label{its_rl}
\end{figure}

\indent The first data-driven research opportunity worth exploring is the use of reinforcement learning (RL) to perform dynamic ITS orchestration. Based on the dynamic nature of ITSs, particularly with the high mobility characterizing some of the entities (e.g. vehicles, VRUs, etc.), data-driven intelligent algorithms capable of adapting to this dynamic nature are required. Accordingly, advanced ML paradigm such as RL become a viable potential solution  \cite{RL_networks}. More specifically, RL algorithms can learn optimal policies for ITS orchestration and apply them. Additionally, such algorithms can meet the stringent performance requirements of ITS services since they typically have lower run-time complexity than the traditional ML algorithms.\\
\indent Fig. \ref{its_rl} illustrates a potential RL-enabed ITS orchestration architecture. More specifically, the RL learning agent would gather data from the different involved entities such as the vehicles, RSUs, VRUs, and cellular base station. This data is then used to learn optimal orchestration policies such as channel conditions or user densities. The ITS can then implement these policies during future decision making processes without incurring high computational complexity.
\subsection{ITS Traffic \& Channel Condition Prediction Using Supervised Learning:}
\begin{figure}[!htb]
	\centering
	\includegraphics[scale=.32,trim=0cm 0cm 0cm 1.5cm]{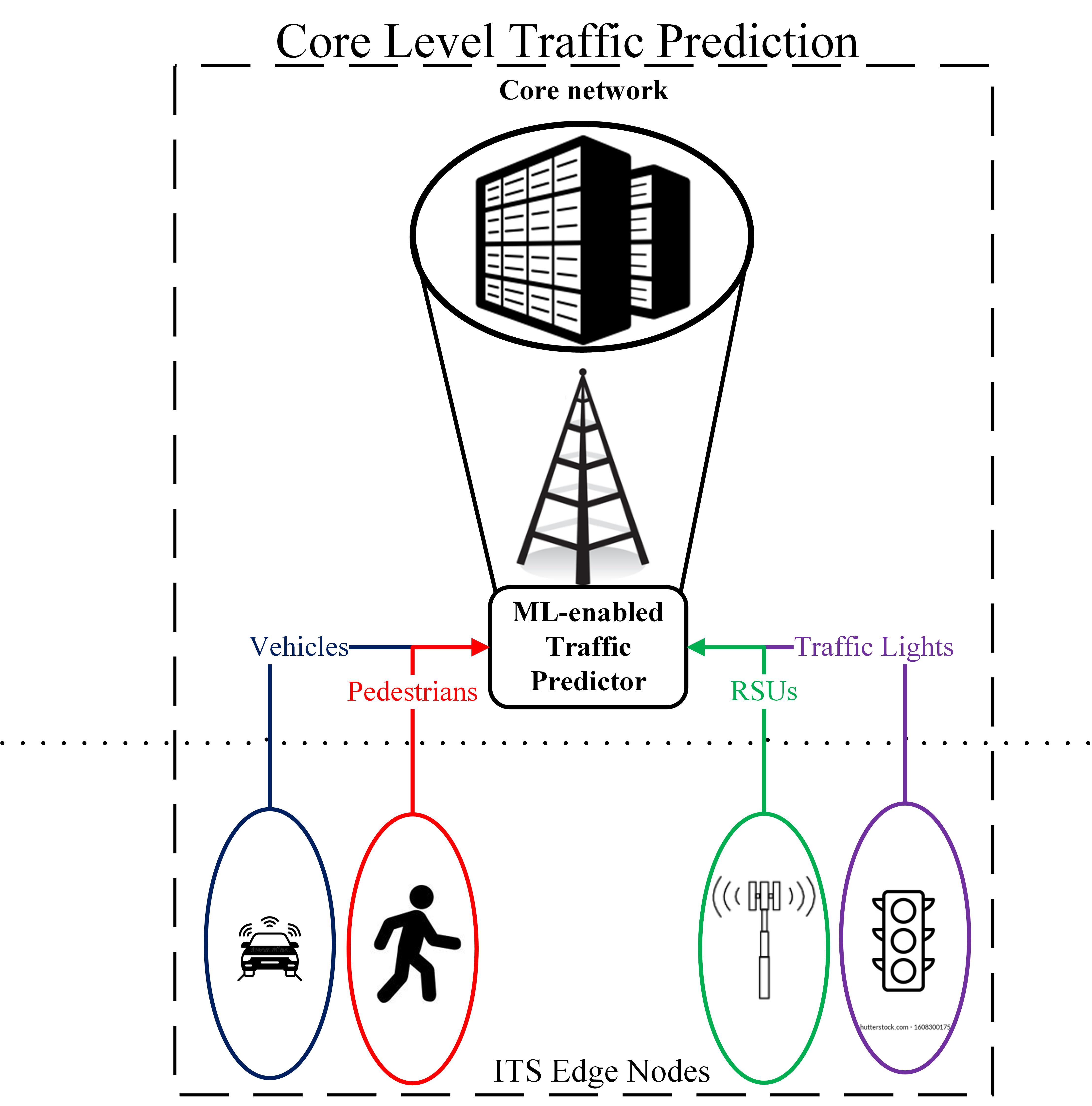}
	\caption{Potential ML-enabled ITS Traffic Prediction Architecture}
	\label{its_ML_traffic_predict}
\end{figure}

\indent One important aspect that can facilitate the orchestration process is having an accurate prediction of the incoming ITS traffic. This is because the incoming traffic and its performance requirements has a direct impact on the ITS orchestration decisions made. Accordingly, supervised learning algorithms can be used to predict the time, volume, and type of incoming traffic. Using this traffic prediction, service providers can plan ahead of time by scaling up or down the available resources (communication and computing) and make service provisioning decisions ahead of time with the goal of satisfying the incoming traffic's requirements. Moreover, such processes become vital in understanding the traffic characteristics, especially with the constantly changing user behaviour.\\
\indent Fig. \ref{its_ML_traffic_predict} is an illustrative example of how to incorporate supervised ML algorithms to achieve accurate ITS traffic prediction. More specifically, the ML traffic prediction module can be deployed at the cellular base stations supporting the ITSs. These base stations then would collect end-user/vehicle data directly (from the end-user/vehicle to the base station) or indirectly (from end-user/vehicle to RSUs and traffic lights then to base station). This data can then be shared with the core network to make effective orchestration decisions in terms of allocating communication resources and provisioning the corresponding ITS services on the available computing resources.\\

\indent Another important aspect that can directly impact the orchestration decisions is the channel conditions. In this case, service providers would use the data collected at the base stations and RSUs to deploy supervised learning algorithms such as polynomial regression or support vector regression (SVR) in order to better predict the channel conditions. In turn, these channel predictions along with other features (e.g. access technology congestion level, application/service performance requirements, etc.) can then be fed into another supervised learning algorithms such as support vector machines and random forests to select the appropriate access technology for each vehicle and application/service demand. Therefore, combining ITS traffic and channel conditions prediction can facilitate the orchestration process.
\subsection{Distributed ITS Orchestration Using Federated Learning:}
\begin{figure}[!htb]
	\centering
	\includegraphics[scale=.32,trim=0cm 0.5cm 0cm 0.5cm]{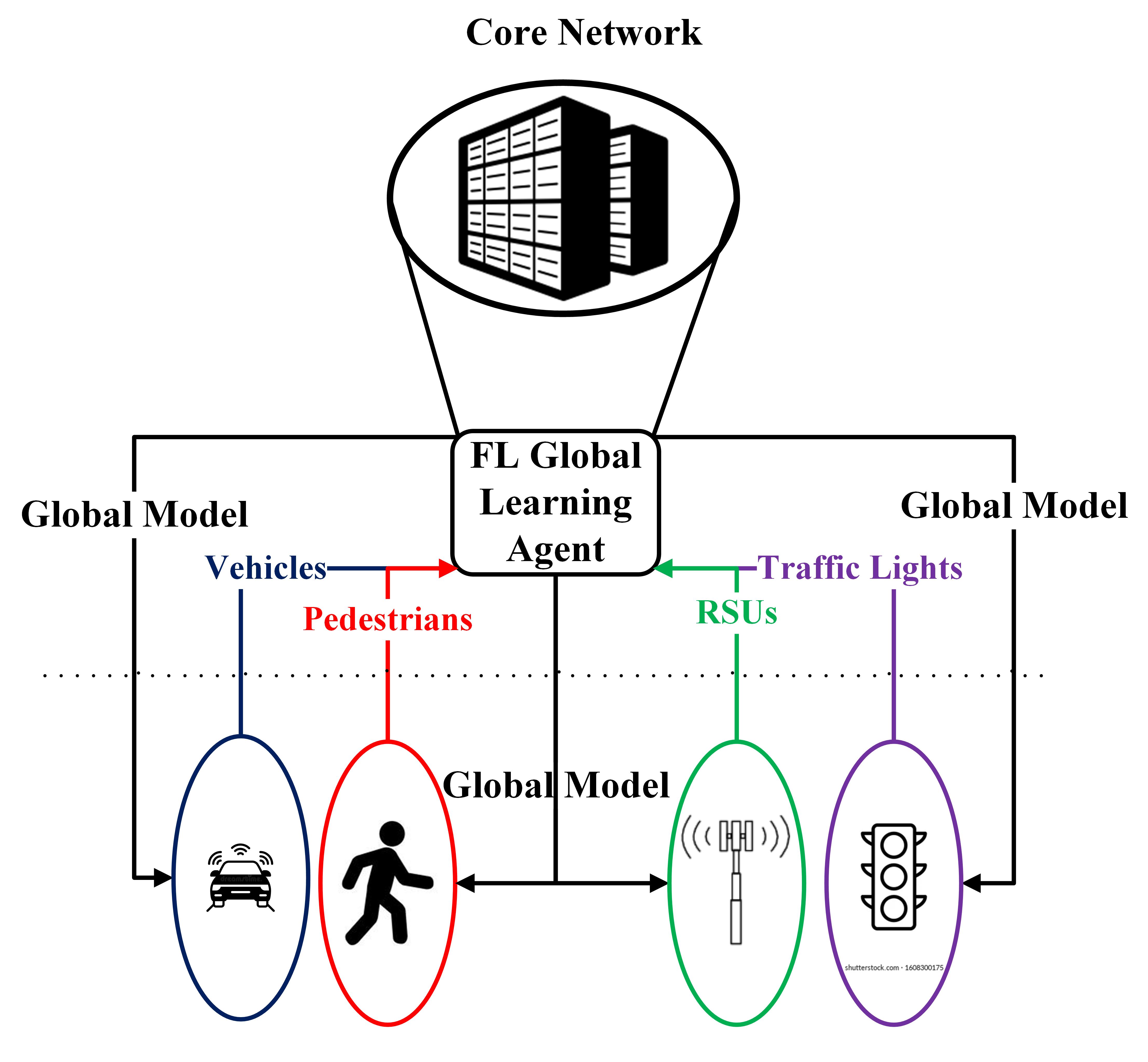}
	\caption{Potential FL-enabled ITS Orchestration Architecture}
	\label{its_fl}
\end{figure}

\indent Another potential data-driven research opportunity is using federated learning (FL) to achieve distributed ITS orchestration and decision making. The importance of distributed orchestration and decision making is highlighted by the fact that ITSs constantly change in size and complexity as well as use a variety of enabling technologies. Hence, there is a need for intelligent modules that can benefit from the multitude of data collected. Accordingly, FL can be adopted in this case since it allows different agents to cooperate and collaborate without them having to share any sensitive information \cite{FL_moubayed}. As a result, the insights gained from multiple locations and technologies is shared without the complexity and security aspects being a concern.\\
\indent Fig. \ref{its_fl} presents a potential FL-enabled architecture for ITS orchestration. The local learning agents in this case would collect data from the different entities within the ITS and share their individual model parameters with the global learning agent. In turn, the global learning agent would use these local models to generate a global model that is then shared back with the local agents. Accordingly, better local orchestration decisions can be taken using the insights gained from other locations/entities without sacrificing the security/privacy of these entities. 

\section{Conclusion}\label{conc}
\indent The continually growing deployment efforts of 5G networks globally has led to the acceleration of the businesses/services' digital transformation. This growth has led to the need for new communication technologies that will promote this transformation while also enabling the sustainability of the systems available. 6G is being proposed as the set of technologies and architectures that will achieve this target. Among the main use cases that have emerged for 5G networks and will continue to play a pivotal role in 6G networks is that of Intelligent Transportation Systems (ITSs). The development of efficient ITSs has attracted attention from a multitude of stakeholders due to the various benefits it offers. \\
\indent With all the projected benefits of developing and deploying efficient and effective ITSs comes a group of unique challenges that need to be addressed. One prominent challenge is ITS orchestration due to the various supporting technologies (e.g. softwarization, virtualization, etc.) and heterogeneous networks used to offer the desired ITS applications and services. Hence, managing and coordinating these networks by allocating the appropriate available resources is challenging. \\
\indent To that end, this paper focused on the ITS orchestration challenge in detail by highlighting the related previous works from the literature and listing the lessons learned from current ITS deployment orchestration efforts. It also presented three potential data-driven research opportunities in which paradigms such as RL and FL can be deployed to offer effective and efficient ITS orchestration.

\bibliographystyle{IEEEtran}
\bibliography{Bibliography}

\end{document}